\documentclass[aps,12pt,showpacs,preprintnumbers,amsmath,amssymb]
{revtex4}\draft
\usepackage[dvips]{graphicx}
\usepackage{bm}
\begin{document}
\title{Construction and enlargement of traversable wormholes from
Schwarzschild black holes}
\author{Hiroko Koyama}
\email{koyama@gravity.phys.waseda.ac.jp}
\affiliation{Department of Physics, Waseda University, Shinjuku, Tokyo 169-8555, Japan}
\author{Sean A. Hayward}
\email{sean_a_hayward@yahoo.co.uk} \affiliation{Department of Science
Education, Ewha Womans University, Seoul 120-750, Korea}
\date{revised 27th July 2004}

\begin{abstract}
Analytic solutions are presented which describe the construction of a
traversable wormhole from a Schwarzschild black hole, and the enlargement of
such a wormhole, in Einstein gravity. The matter model is pure radiation which
may have negative energy density (phantom or ghost radiation) and the
idealization of impulsive radiation (infinitesimally thin null shells) is
employed.
\end{abstract}
\pacs{04.20.Jb, 04.70.Bw} \maketitle

\section{Introduction}
While black holes are now almost universally accepted as astrophysical
realities, traversable wormholes are still a theoretical idea \cite{MT,V}. Yet
they are both predictions of General Relativity in a sense, though black holes
require positive-energy matter (or vacuum) whereas wormholes require
negative-energy matter. While normal positive-energy matter was long thought to
dominate the universe, it is now known that this is not so. The recently
discovered acceleration of the universe \cite{Spe,Kra} indicates that its
evolution is dominated by unknown dark energy which violates at least the
strong energy condition ($w\ge-1/3$ to cosmologists, where $w$ is the ratio of
pressure to density in relativistic units, for a homogeneous isotropic cosmos),
and perhaps also the weak energy condition ($w\ge-1$), where it is known as
phantom energy \cite{Cal}. Such phantom energy is precisely what is needed to
support traversable wormholes \cite{HV1,HV2,wh,IH}.

While black holes and traversable wormholes have been regarded by most experts
as quite different, one of the authors has argued that they form a continuum
and are theoretically interconvertible \cite{wh}. Specifically, both are
locally characterized by trapping horizons \cite{wh,IH,1st,bhd}, which are the
Killing horizons of a stationary black hole and the throat of a stationary
wormhole. The difference is the causal nature, being spatial or null for a
black hole and temporal for a wormhole. This in turn depends on whether the
energy density is positive, zero or negative. If the energy density can be
controlled, it should be possible to dynamically create a traversable wormhole
from a black hole and vice versa. This was first concretely demonstrated in a
two-dimensional model \cite{HKL}, but it is more difficult in full General
Relativity. Numerical simulations have been used to study a wormhole supported
by a ghost (or phantom) scalar field, showing that it does indeed collapse to a
black hole if perturbed by positive-energy matter \cite{SH}. (We use phantom to
mean that the energy density has the opposite sign to normal, which is
equivalent in our cases to the convention for ghost fields in quantum field
theory, that the kinetic energy has the opposite sign). As for analytic
results, one simple case is a static wormhole supported by pure ghost (or
phantom) radiation \cite{pr}; it is easy to see that if the radiation is
switched off, it immediately collapses to a Schwarzschild black hole. The
converse, creating a traversable wormhole from a Schwarzschild black hole, is
more complex and is the first main result of this article.

As above, we use pure phantom radiation as the exotic matter model. We also
employ the idealization of impulsive radiation, where the radiation forms an
infinitesimally thin null shell, thereby delivering finite energy-momentum in
an instant \cite{KHK}. Space-time regions can be matched across such shells
using the Barrab\`es-Israel formalism \cite{B-I}. This allows an ingenious
analytic construction of the desired type of solution, by matching
Schwarzschild, static-wormhole and Vaidya regions, the latter consisting of
pure radiation propagating in a fixed direction \cite{Vaidya}. Our other main
result is the similar construction of analytic solutions describing the
enlargement or reduction of such a wormhole. Then if Wheeler's space-time foam
picture \cite{Whe} is correct and Planck-sized virtual black holes are
continually forming, we have exact solutions in standard Einstein gravity
describing how they may be converted into traversable wormholes and enlarged to
usable size. Our results have been summarized in a shorter article \cite{HK}.

Our paper is organized as follows. In Sec.~\ref{sec:model}, we briefly review
the static wormhole solutions with pure phantom radiation, and the
Schwarzschild and Vaidya solutions. In Sec.~\ref{sec:matching} we show how to
join these basic solutions at null boundaries. In Secs.~\ref{sec:const} and
\ref{sec:enlarge} we present analytic solutions which describe respectively the
construction and enlargement of a wormhole. In Sec.~\ref{sec:jump} we consider
the jump in energy due to impulsive radiation, as a check that the matchings
are physically reasonable, and as a simple way to understand the changes in
area of the wormhole throat or black-hole horizon. The final section is devoted
to summary.

\section{Basic solutions}
\label{sec:model} In this section we review the traversable wormhole solution
\cite{pr}, the Schwarzschild solution and the Vaidya solutions in the various
coordinates needed. We will consider spherically symmetric space-times only. It
is convenient to use the area radius $r=\sqrt{A/4\pi}$, where $A$ is the area
of the spheres of symmetry. Although we often use $r$ as a coordinate, it is a
geometrical invariant of the metric and is assumed throughout to be continuous.
A useful quantity is the local gravitational mass-energy \cite{MS,sph}
\begin{equation}\label{energy}
E=\frac{r}{2}(1-g^{\mu\nu}r_{,\mu}r_{,\nu}).\end{equation} Note that $E=r/2$ on
a trapping horizon, where $g^{\mu\nu}r_{,\mu}r_{,\nu}=0$, including both
black-hole horizons \cite{bhd} and wormhole mouths \cite{wh}. The
energy-momentum tensor of pure radiation (or null dust) is $T_{ab}=\rho
u_au_b$, where $u_a$ is null and $\rho$ is the energy density. Normally
$\rho\ge0$, but $\rho<0$ defines pure phantom radiation.

\subsection{Static wormhole solution with pure phantom radiation}
\label{subsec:wormhole} The static wormhole solutions \cite{pr} supported by
opposing streams of pure phantom radiation can be written as
\begin{equation}
ds^2=-\frac{2\lambda}{1+2l\phi e^{l^2}}dt^2+\frac{1+2l\phi e^{l^2}}{2\phi^2e^{2l^2}}dr^2+r^2d\Omega^2
\label{metric:wh1}\end{equation}
where $t$ is the static time coordinate and $d\Omega^2$ refers to the unit
sphere. Here $l$ is a function of $r$,
\begin{equation}
\label{r}
 r= a(e^{-l^2}+2l\phi),
\end{equation}
and $\phi$ is an error function,
\begin{equation}
\label{phi} \phi (l)\equiv  \int _{0}^{l} e^{-\ell^2}d\ell +b,
\end{equation}
where $a>0$, $b$ and $\lambda>0$  are constants.

The local energy evaluates as $E=\epsilon$ where
\begin{equation}
\epsilon =\frac{a}{2}(e^{-{l}^{2}}+2l\phi -2e^{{l}^{2}}\phi^2).
\end{equation}
Using the mass-energy $\epsilon$, the metric (\ref{metric:wh1}) is rewritten as
\begin{equation}
\label{metric:wh}
ds^2=- \frac{\lambda}{e^{2l^2}\phi^2}
\left(1-\frac{2\epsilon}{r}\right)dt^2+
\left(1-\frac{2\epsilon}{r}\right)^{-1}dr^2+r^2d\Omega ^2.
\end{equation}
In the case $b=0$, the solutions (\ref{metric:wh1}) or (\ref{metric:wh})
describe symmetric wormholes. The spacetime is not asymptotically flat, but
otherwise constitutes a Morris-Thorne wormhole. The $b\ne 0$ cases include
asymmetric wormholes which are analogous to the asymmetric Ellis wormhole for a
phantom Klein-Gordon field \cite{Ellis}. For the space-time solution to be a
wormhole, the inequality
\begin{equation}
|b|<b_{cr}=\frac{\sqrt{\pi}}{2}
\end{equation}
is needed. For any other value of $b$, a singularity is present. Hereafter, we
consider $b=0$, describing a symmetric wormhole with minimal surfaces at the
wormhole throat $l=0$, with area radius $r=a$.

\begin{figure}[t]
\includegraphics[height=4cm]{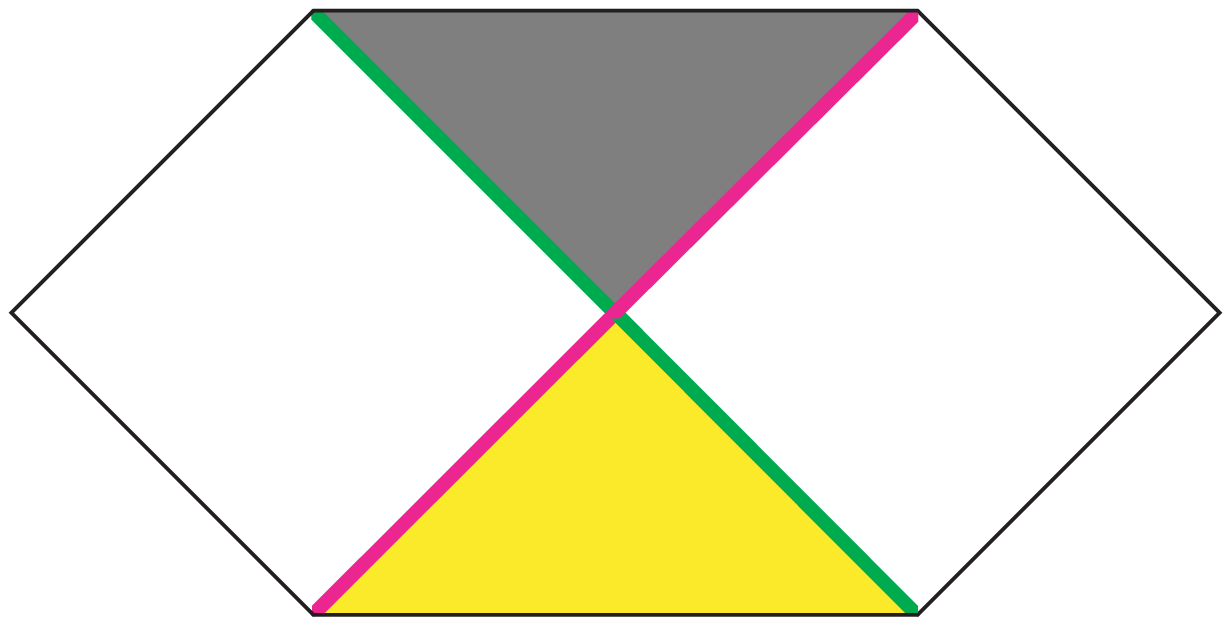}\hspace*{1cm}
\includegraphics[height=5cm]{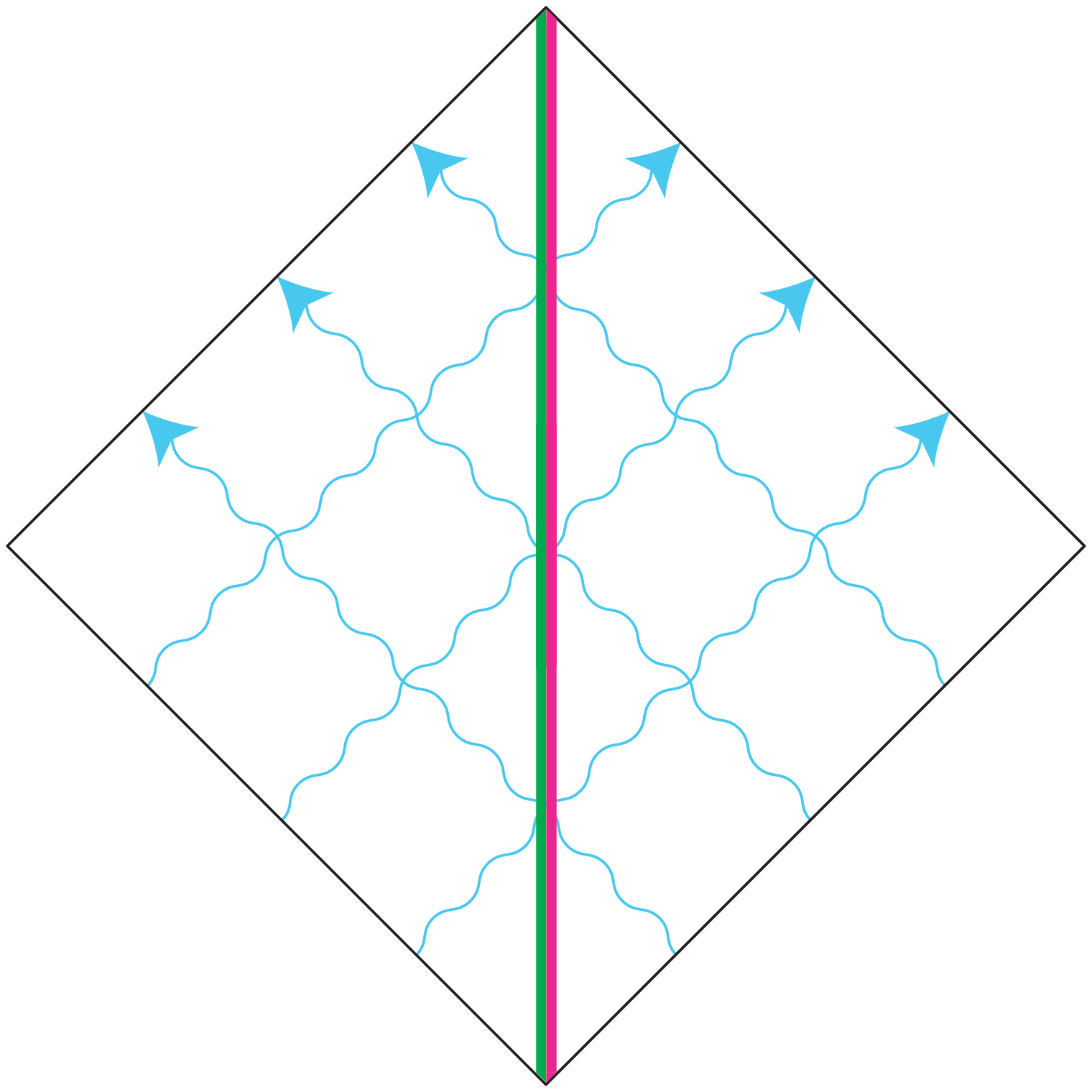}
 \caption{(Color online). Penrose diagrams of (i) a Schwarzschild black hole and (ii)
a Hayward traversable wormhole \cite{pr}. The bold magenta and green lines
represent the trapping horizons, $\partial_{+}r=0$ and $\partial_{-}r=0$,
respectively, which constitute the event horizons of the black hole and the
throat of the wormhole. Yellow (light) and gray (dark) quadrants represent past
trapped and future trapped regions, respectively. Wavy cyan lines represent the
constant-profile radiation supporting the wormhole structure.}
 \label{fig:bh-wh}
\end{figure}

The solutions (\ref{metric:wh}) may be written in dual-null form
\begin{equation}\label{dual-null}
 ds^2=-\frac{2\lambda}{1+2le^{l^2}\phi}dx^+dx^-+r^2d\Omega^2
\end{equation}
where the null coordinates $x^{\pm}$ are defined by
\begin{equation}\label{xpm}
  dx^{\pm}=dt\pm \frac{a}{\sqrt{\lambda}}\left(e^{-l^2}+2l\phi \right)dl.
\end{equation}
Then the radial null geodesics are given by constant $x^{\pm}$. We need the
metric (\ref{metric:wh}) in both ingoing and outgoing radiation coordinates:
\begin{equation}
\label{metric:eddinton}
ds^2=-\frac{\sqrt{\lambda}}{e^{l^2}\phi}dx^{\pm}
\left[\frac{2\sqrt{\lambda}e^{l^2}\phi}{(1+2le^{l^2}\phi)}dx^{\pm}\mp 2 dr\right]+r^2d\Omega^2.
\end{equation}
In these coordinates the radial null geodesics are the lines of constant
$x^\pm$ (choosing one) and the curves given by
\begin{equation}
\label{eddinton-geodesics}
\frac{dr}{dx^{\pm}}=\pm\frac{\sqrt{\lambda}e^{l^2}\phi}{(1+2le^{l^2}\phi)}
.\end{equation} The Penrose diagram is shown by Fig.\ref{fig:bh-wh} (ii). The
energy-momentum tensor supporting the wormhole is found to be
\begin{equation}
\label{whem} T_{ab}= -\frac\lambda{8\pi
r^2}(\delta^{+}_a\delta^{+}_b+\delta^{-}_a\delta^{-}_b).
\end{equation}
This is the energy tensor of two opposing streams of pure phantom radiation,
with $\lambda=-4\pi r^2T_{tt}$ being the resulting negative linear energy
density. On the other hand, the solutions
with pure radiation of the usual positive energy density were founded by Gergely \cite{Gergely}.

\subsection{Schwarzschild solution}
The Schwarzschild metric is given by
\begin{equation}
\label{metric:Schwarzschild}
ds^{2}= - \left(1-\frac{2M}{r}\right)dt^2
+\left(1-\frac{2M}{r}\right)^{-1}dr^2
+ r^{2} d\Omega^2,
\end{equation}
where the constant $M$ is the Schwarzschild mass, which coincides with the
local energy, $E=M$. Rewriting in Eddington-Finkelstein
coordinates
\begin{equation}\label{EF} V=t-\zeta\left(r+2M\ln(1-r/2M)\right)
\end{equation}
one finds
\begin{equation}
\label{metric:Schwarzschild-Eddington}
ds^{2}= - dV \left[\left(1-\frac{2M}{r}\right)dV +2 \zeta dr\right]
+ r^{2} d\Omega^2,
\end{equation}
where $\zeta$ is a sign factor: $\zeta$ is $1$ for outgoing radiation or $-1$
for ingoing radiation, where this means that the area respectively increases or
decreases along the future-null generators. The Penrose diagram is shown by
Fig.\ref{fig:bh-wh} (i).

\subsection{Vaidya solutions}
The metric of the Vaidya solutions is given by
\begin{equation}
\label{metric:Vaidya}
ds^{2}= - dV \left[\left(1-\frac{2m(V)}{r}\right)dV +2 \zeta dr\right]
+ r^{2} d\Omega^2,
\end{equation}
where $\zeta$ is a sign factor, where $\zeta =1$  for outgoing radiation, and
$\zeta =-1$ for ingoing radiation. The mass function $m$ coincides with the
local energy, $E=m$. The corresponding energy-momentum tensor is given by
\begin{equation}
\label{eq14} T_{\mu\nu} = - \frac{\zeta}{4\pi r^{2}}\frac{dm}{dV} \delta^{V}_{\mu}
\delta^{V}_{\nu}.
\end{equation}

\section{Matching Vaidya regions to static-wormhole and Schwarzschild
 regions}
\label{sec:matching} In this section, we derive the matching formulas between
Schwarzschild, Vaidya and static-wormhole regions along null hypersurfaces,
following the Barrab\`es-Israel formalism \cite{B-I}.
This is a preliminary to constructing the wormhole-construction and
wormhole-enlargement models.

\subsection{Matching Vaidya and Schwarzschild regions}
Firstly we consider the matching between Schwarzschild and Vaidya regions. We
start by writing Schwarzschild and Vaidya solutions in the form
\begin{equation}
\label{eq12}
ds^{2}= - e^{\psi}dV (f e^{\psi}dV + 2\zeta dr)
+ r^{2} d\Omega^2,
\end{equation}
where the metric functions are
\begin{equation}
\label{def_Sch}
f_{S}=1-\frac{2M}{r},\qquad \psi_S=0
\end{equation}
for Schwarzschild, and
\begin{equation}
\label{def_Vaidya}
f_{V}=1-\frac{2m(V)}{r},\qquad \psi_V=0
\end{equation}
for Vaidya.

\begin{figure}[h]
\begin{center}
\includegraphics[width=5cm]{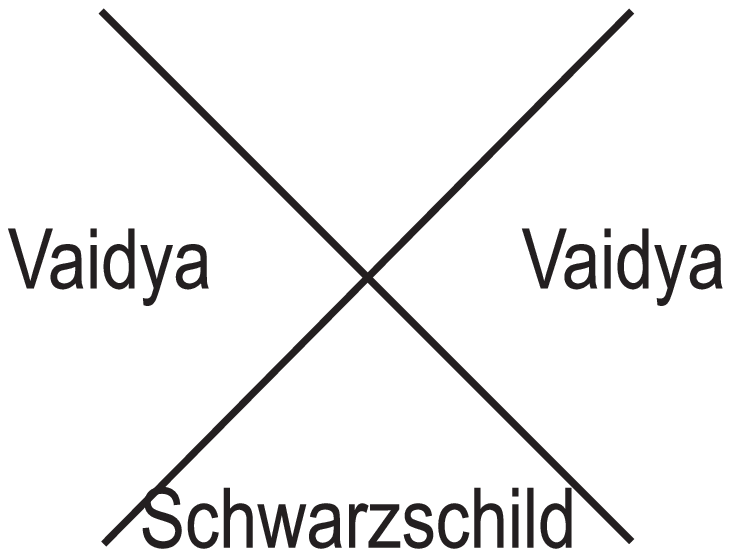}
\hspace{4cm}
\includegraphics[width=5cm]{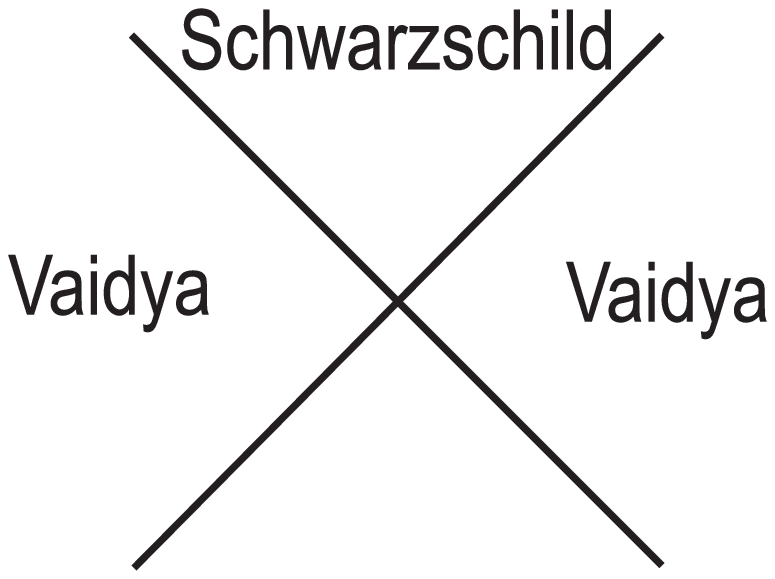}
\caption{Matching domains: $\eta_1=1$ (left) and $\eta_1=-1$ (right). }
\label{fig:matchSV}
\end{center}
\end{figure}

Now we consider the boundary surface $V=V_0$ (constant). The normal to the
hypersurface $\Phi = V- V_{0}= 0$ is $n_{\mu} =
\zeta\alpha^{-1}\partial_{\mu}\Phi =\zeta\alpha^{-1}\delta^{V}_{\mu}$, where
$\alpha$ is a positive function. (Barrab\`es \& Israel took $\alpha<0$). For a
null hypersurface, the normal is also tangent, so to obtain extrinsic curvature
one needs a different vector. From $n_{\mu}$ Barrab\`es and Israel introduced a
so-called transverse null vector $N_{\mu}$ by requiring $N_{\mu}N^{ \mu} = 0$,
and $N_{\mu}n^{\mu} = - 1$. Without loss of generality, we assume that
$N_{\mu}$ takes the form $N_{\mu} =N_{V}\delta^{V}_{\mu} +
N_{r}\delta^{r}_{\mu}$, and choose the arbitrary function $\alpha$ as $\alpha
=e^{-\psi}$. Then $N_{\mu}$ is given by
\begin{equation}
\label{eq8}
N_{\mu} = \zeta \frac{f e^{\psi}}{2}\delta^{V}_{\mu} + \delta^{r}_{\mu}.
\end{equation}
Choosing the coordinates $r, \theta$, and $\varphi$ as the three
intrinsic coordinates $\xi^{a} \equiv (r, \theta, \varphi), \; (a = 1,
2, 3)$ on the hypersurface $V = V_{0}$, we find
\begin{equation}
\label{eq9}
e^{\mu}_{(1)} = \delta^{\mu}_{r},\;\;\;
e^{\mu}_{(2)} = \delta^{\mu}_{\theta},\;\;\;
e^{\mu}_{(3)} = \delta^{\mu}_{\varphi},
\end{equation}
where $e^{\mu}_{(a)} \equiv \partial x^{\mu}/\partial\xi^{a}$. Then,
it can be
shown that the transverse extrinsic curvature, defined by \cite{B-I}
\begin{equation}
\label{eq10}
{\cal{R}}_{a b} = - N_{\mu} e^{\nu}_{(b)}
\left(\nabla_{\nu}e^{\mu}_{(a)}\right),
\end{equation}
takes the form
\begin{equation}
\label{eq11}
{\cal{R}}_{a b} = diag.
\left\{ \zeta\frac{\partial \psi}{\partial r},
\; -\zeta\frac{r f}{ 2},\; -\zeta\frac{r f}{2}\sin^{2}\theta\right\}.
\end{equation}
The jump in transverse extrinsic curvature is denoted by
\begin{equation}
\label{gamma}
\gamma_{a b}=2 [{\cal{R}}_{a b}].
\end{equation}
Once $\gamma_{a b}$ is given, using the formula \cite{B-I}
\begin{equation}
\label{eq21} \tau^{a b} = - S^{a b} = \frac{1}{16 \pi}(g^{a c}_{*} l^{b}l^{d} +
g^{b d}_{*} l^{a}l^{c} - g^{a b}_{*} l^{c}l^{d} - g^{c d}_{*}
l^{a}l^{b})\gamma_{cd},
\end{equation}
we can calculate the surface energy-momentum tensor $\tau^{a b}$ on the null
hypersurface $V = V_{0}$. In components,
\begin{equation}
\label{eq22}
\tau^{a b} = \sigma l^{a}l^{b} + P g^{a b}_{*},
\end{equation}
where
\begin{eqnarray}
\label{eq24} g^{a b}_{*} &=& r^{-
2}\left(\delta^{a}_{\theta}\delta^{b}_{\theta}
+ \sin^{- 2}\theta\delta^{a}_{\varphi }\delta^{b}_{\varphi}\right),\nonumber \\
&& l^{a} = \delta^{a}_{r},\;\;\;\;\;\; l^{b}l_{b} = 0.
\end{eqnarray}
Here $\sigma$ represents the surface energy density of the null shell and $P$
the pressure in the $\theta$ and $\varphi$ directions. Then
\begin{eqnarray}
\label{matchVS}
\sigma &=& \zeta\frac{[f]}{8\pi r}=
\eta_1\zeta\frac{M-m(V_0)}{4\pi r^2},\nonumber\\
P &=&  -\frac{\zeta}{8\pi} \left[\frac{\partial\psi}{\partial r}\right]=0.
\end{eqnarray}
We take the sign factor $\eta_1$ to be $1$ if the radiation is to the future of
the Schwarzschild region, and $-1$ if the radiation is to the past of the
Schwarzschild region (Fig.\ref{fig:matchSV}).
The energy-momentum tensor of the impulsive radiation is
given generally by \cite{B-I}
\begin{equation}
T^{\mu\nu}=\alpha\tau^{\mu\nu}\delta(\Phi),
\end{equation}
where $\delta$ denotes the Dirac delta distribution. In our case it reduces to
\begin{equation}
\label{ET-sv}
T^{rr}=\tau ^{rr} \delta(V-V_0)
=\eta_1\zeta\frac{[M-m(V_0)]}{4\pi r^2}\delta(V-V_0).
\end{equation}
\subsection{Matching Vaidya and static-wormhole regions}
\begin{figure}[h]
\begin{center}
\includegraphics[width=5cm]{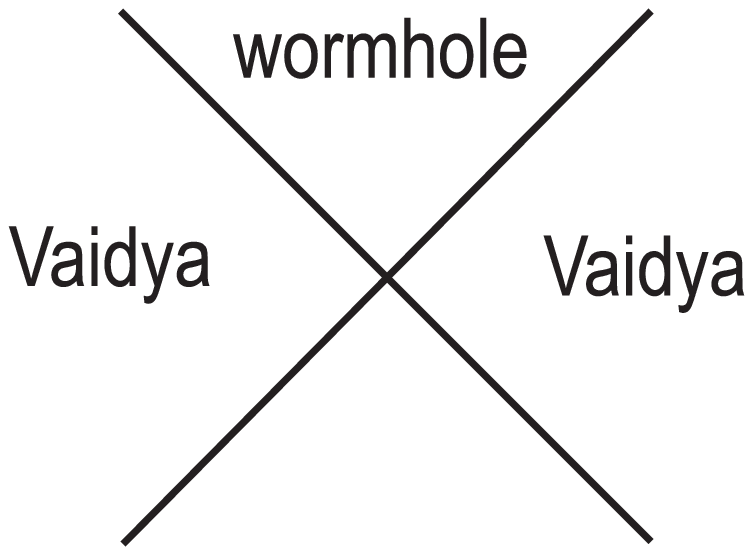}
\hspace{4cm}
\includegraphics[width=5cm]{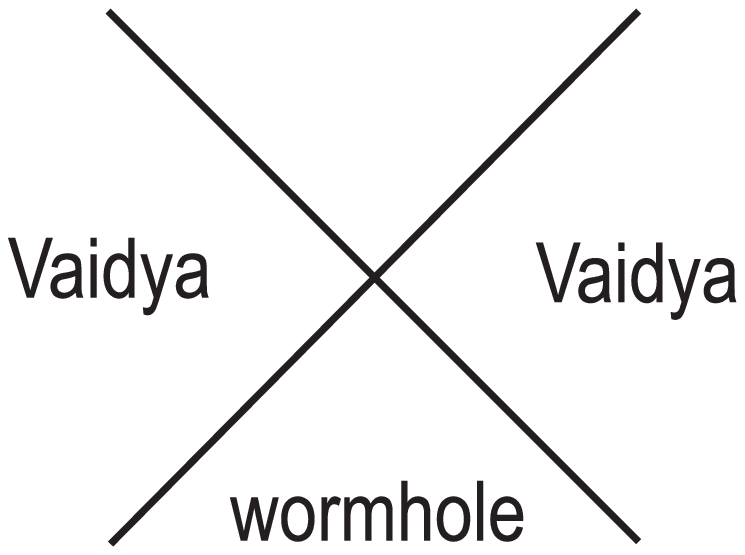}
\caption{Matching domains: $\eta_2=1$ (left) and $\eta_2=-1$ (right). }
\label{fig:matchVW}
\end{center}
\end{figure}

Secondly we consider matching Vaidya and static-wormhole regions. We rewrite
the static wormhole solution (\ref{metric:eddinton}) as
\begin{equation}
\label{metric_B-I}
ds^2=-e^{\psi}du(f_We^{\psi_W}du-2\zeta dr)+r^2d\Omega^2,
\end{equation}
where $u$ is $x^{\pm}$ for $\zeta =\pm 1$. The metric functions $f_W$ and
$\psi_W$ are defined as
\begin{equation}
\label{}
f_W=1-\frac{2\epsilon}{r}, \qquad
e^{\psi_W}=\frac{\sqrt{\lambda}}{e^{l^2}\phi}.
\end{equation}
Now we consider the boundary surface $u=u_0$ (constant). Then, from the
previous subsection, the transverse extrinsic curvature takes the form
\begin{equation}
\label{eq11}
{\cal{R}}^{+}_{a b} = diag.\left\{-\zeta\frac{\partial \psi_{W}}{\partial r},
\; \zeta\frac{r f_{W}}{2},\; \zeta\frac{r f_{W}}{2}\sin^{2}\theta\right\}.
\end{equation}
On the other hand, we write the Vaidya solution \cite{Vaidya} in the same form
as (\ref{def_Vaidya})
\begin{equation}
\label{eq12}
ds^{2} = - e^{\psi_{V}}dV (f_{V} e^{\psi_{V}}dV + 2\zeta  dr)
+ r^{2} d\Omega^2,
\end{equation}
where the metric $f_{V}$ and $\psi_{V}$ is defined by (\ref{def_Vaidya}).
The hypersurface $u = u_{0}$ in the $(V,r)$ coordinates
can be written as $\Phi^{-} = V - V_{0}(r) = 0$, where
$V_{0}(r)$ is a solution of the equation
\begin{equation}
\label{eq16}
\frac{dV_{0}}{dr} = -\zeta\frac{2 }{f_{V}} e^{- \psi_{V}}, \;\;\; (u = u_{0}).
\end{equation}
Then the normal to the surface is given by
\begin{equation}
n^{-}_{\mu}
=\zeta\beta^{- 1}\partial_{\mu}\Phi^{-}
=\zeta\beta^{- 1}\left(\delta^{V}_{\mu} +\frac{2}{f_{V}}e^{ - \psi_{V}}\delta^{r}_{\mu}\right),
\end{equation}
where $\beta$ is a negative and otherwise arbitrary function. From
$n^{-}_{\mu}$ we can also introduce the transverse null vector $N^{-}_{\mu}$,
by requiring $N^{-}_{\mu}N^{- \mu} = 0$, and $N^{-}_{\mu}n^{- \mu} = - 1$. It
can be shown that it takes the form
\begin{equation}
\label{eq17}
N^{-}_{\mu} = \zeta\frac{\beta f_{V}}{2} e^{2\psi_{V}}\delta^{V}_{\mu}.
\end{equation}
The basis vectors are
\begin{equation}
\label{eq18}
e^{- \mu}_{(1)}
= -\zeta \frac{2}{f_{V}}e^{- \psi_{V}}\delta^{\mu}_{V}
+ \delta^{\mu}_{r},\;\;\;
e^{- \mu}_{(2)} = \delta^{\mu}_{\theta},\;\;\;
e^{- \mu}_{(3)} = \delta^{\mu}_{\varphi},
\end{equation}
where $e^{- \mu}_{(a)} \equiv \partial x^{\mu}_{-}/\partial\xi^{a}$. To be sure
that the two transverse vectors $N^{\pm}_{\mu}$ defined in the two faces of the
hypersurface $u = u_{0}$ represent the same vector, we need to impose the
condition
\begin{equation}
N^{+}_{\lambda} e^{+ \lambda}_{(a)}|_{u = u_{0}}
= N^{-}_{\lambda} e^{- \lambda}_{(a)}|_{u = u_{0}},
\end{equation}
which requires that the function $\beta$ has to be $\beta = -
\exp\{-\psi_{-}\}$. Once $N^{-}_{\lambda}$ and $e^{- \lambda}_{(a)}$ are given,
using Eq.(\ref{eq10}) we can calculate the corresponding transverse extrinsic
curvature, which in the present case takes the form
\begin{equation}
\label{eq19}
{\cal{R}}^{-}_{a b} = diag.\left\{ -\zeta
\frac{2e^{- \psi_{V}}}{f_{V}^{2}}\frac{\partial f_{V}}{\partial V},\;
\zeta\frac{r f_{V}}{2},\;
\zeta\frac{r f_{V}}{2} \sin^{2}\theta\right\}.
\end{equation}
Then, from Eqs.(\ref{eq19}) and (\ref{eq11}), we find the surface
energy-momentum tensor (\ref{eq21}) on the null hypersurface $u= u_{0}$,
composed of the surface energy density $\sigma$ of the null shell and the
pressures $P$ in the $\theta$- and $\varphi$-directions (\ref{eq22}), as
\begin{eqnarray}
\label{eq23}
\sigma &=& \eta_2\zeta \frac{f_W-f_V}{8\pi r}=-
\eta_2\zeta \frac{\epsilon(r)-m(r)}{4\pi r^2},\nonumber\\
P &=&  \frac{\eta_2\zeta}{8\pi}\left(
\frac{2e^{- \psi_{V}}}{f_{V}^{2}} \frac{\partial f_{V}}{\partial V}
- \frac{\partial\psi_{W}}{\partial r}\right)
=\frac{\eta_2\zeta}{4\pi}\left[\frac{dm(r)}{dr}
-\frac{m(r)r}{2a^2\phi^2}+\frac{r^2}{4a^2\phi^2}\right],
\end{eqnarray}
where $m(r)$ is the mass function of the Vaidya region on the boundary $u=u_0$,
\begin{equation}
\label{}
m(r)\equiv m(V)|_{u=u_0}.
\end{equation}
Here we take the sign factor $\eta _2$ to be $1$ if the radiation is to the
past of the wormhole region, and to be $-1$ if the radiation is to the future
of the wormhole region (Fig.\ref{fig:matchVW}). In calculating the above
equation, we have not used the particular expressions for the functions $\psi$
and $f$.  Thus, it is valid generally in the case that the boundary surface is
$u=$constant.

Henceforth we consider only the dust shell case $P=0$, then we require
\begin{equation}
\label{eq25}
\frac{dm}{dr}-\frac{mr}{2a^2\phi^2}+\frac{r^2}{4a^2\phi^2}=0.
\end{equation}
Integrating Eq.(\ref{eq25}), we obtain
\begin{equation}
\label{eq:M_V}
m=\frac{a}{2}(e^{-l^2}+2l\phi(l) -2\phi(l)^2e^{l^2})+C\phi(l) e^{l^2},
\end{equation}
where $C$ is an integration constant and related to $\sigma$ by
\begin{equation}
\label{matchVW}
\sigma =\eta_2\zeta C\frac{\phi e^{l^2}}{4\pi r^2}.
\end{equation}
Then if there is no light-like shell, $\sigma=0$ and the mass function is
continuous across the boundary surface, $\epsilon=m$. Extending the relation
(\ref{eq16}) to the Vaidya region, introducing $z(V)$ with $z=l$ on the
boundary surface, we obtain the mass function of the Vaidya solutions beyond
the boundary surface
\begin{equation}
\label{}
m(z)=\frac{a}{2}(e^{-z^2}+2z\phi(z) -2\phi(z)^2e^{z^2})+C\phi(z) e^{z^2}
\end{equation}
where the relation between $z$ and $V$ is
\begin{equation}
\label{V-z}
V(z)=-\zeta\int ^z\frac{2a^2e^{-y^2}(e^{-y^2}+2y\phi(y))}
{a\phi(y)-C}dy.
\end{equation}
Transforming the energy-momentum tensor of the impulsive radiation,
\begin{equation}
\label{ET-wv} T^{rr}=-\tau ^{rr}\delta(u-u_0)=-\tau
^{rr}\delta(V-V_0)\frac{dV}{du} =-\eta_2\zeta C\frac{\sqrt{\lambda}}{4\pi
r^2}\delta(V-V_0).
\end{equation}

\subsection{Combined matching between Schwarzschild and static wormhole via Vaidya}

\begin{figure}[h]
\begin{center}
\includegraphics[width=5cm]{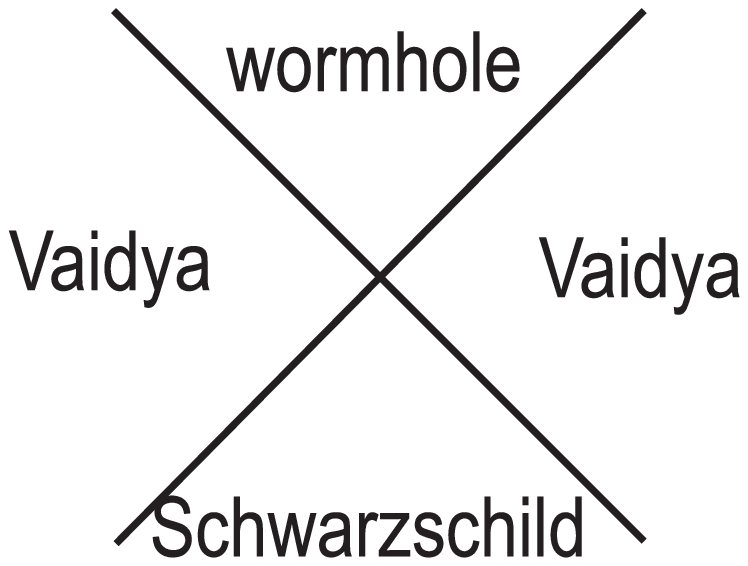}
\hspace{4cm}
\includegraphics[width=5cm]{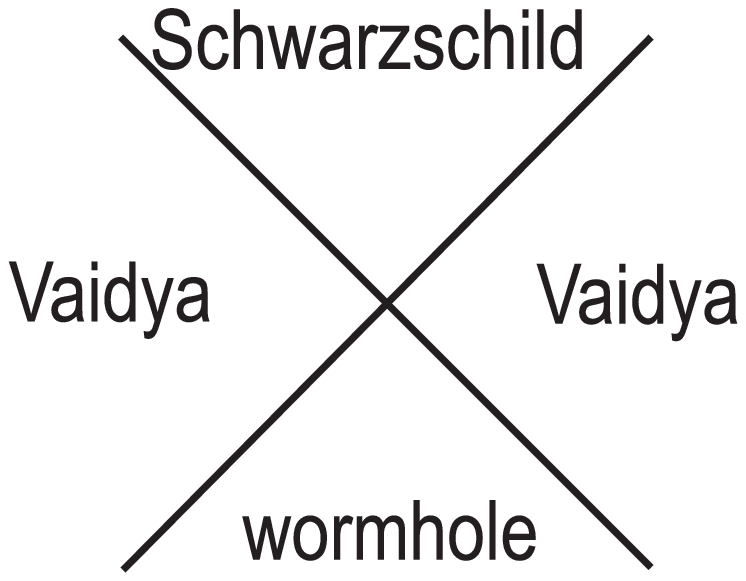}
\caption{Matching domains: $\eta_1=\eta_2=1$ (left) and $\eta_1=\eta_2=-1$
(right). } \label{fig:combinematch}
\end{center}
\end{figure}

In this subsection, we consider a collision of two oppositely moving impulses
with Vaidya regions in opposite quadrants and Schwarzschild and static-wormhole
regions in the other two quadrants. Connecting the impulses by (\ref{ET-sv})
and (\ref{ET-wv}), we find that $\eta_1=\eta_2=1$ if the future region is
wormhole, and $\eta_1=\eta_2=-1$ if the future region is Schwarzschild
(Fig.\ref{fig:combinematch}). That is, $\eta_1=\eta_2$ in both cases. Then the
constant $C$ is determined as
\begin{equation}
\label{eq:c} C=-\frac{M-m(V_0)}{\sqrt{\lambda}}.
\end{equation}
Here we implicitly use the fact that the jump of a jump vanishes, i.e.\ the
jump across one impulse does not jump across the other impulse \cite{Jez}.

\section{Wormhole construction from Schwarzschild black hole}
\label{sec:const} 
\begin{figure}[t]
\begin{center}
\includegraphics[width=10cm]{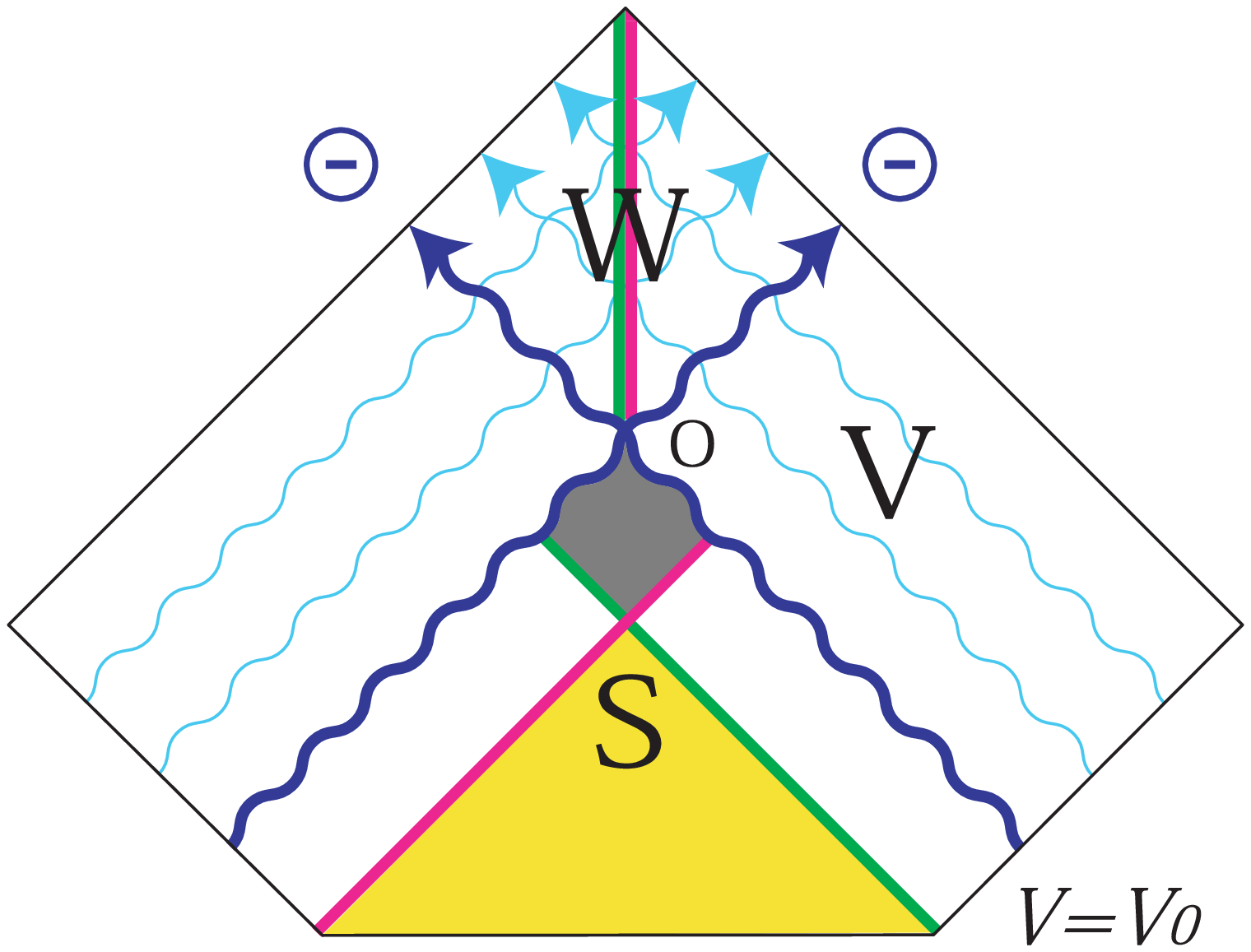}
\caption{(Color online). Penrose diagram of the wormhole construction model.
The wavy blue bold lines represent impulsive radiation with negative energy
density. The region S is Schwarzschild, V is Vaidya and W is static-wormhole.
The boundary between S and V corresponds to $z=0$ and the boundary between V
and W corresponds to $z=l$.} \label{fig:sudden}
\end{center}
\end{figure}

In this section we present analytic solutions which describe
the construction of a static wormhole from a Schwarzschild black hole. The
whole picture is represented by Fig.~\ref{fig:sudden} and the strategy is as
follows. Firstly, impulsive phantom radiation is beamed in, causing the
trapping horizons to jump inward, much as a shell of normal matter makes a
black-hole trapping horizon jump outward. By controlling the energy and timing
of the impulses, the trapping horizons can be made to instantaneously coincide.
They can then form the throat of a static wormhole if constant-profile streams
of phantom radiation are beamed in subsequently, with the energy density
appropriate to a wormhole of that area.

\subsection{Vaidya region V}
First, we set up an initial Schwarzschild region S (\ref{metric:Schwarzschild})
with mass $M$. Now we beam in impulsive phantom radiation symmetrically from
either side, with the mass-energy of the shell being $\mu =4\pi r^2 \sigma $,
then turn on constant streams of phantom radiation immediately after the
impulses. Then the region V should be Vaidya (\ref{metric:Vaidya}) with some
mass function $m$, which on the boundary $V=V_0$ between S and V is
\begin{equation}
\label{eq:m_{II}(V=V_0)} m(V_0)=M-\eta\zeta\mu =M+\mu,
\end{equation}
from the matching formula between Schwarzschild and Vaidya
(\ref{matchVS}).
Here we must take $\zeta =-1$ and $\eta=1$, since the impulse is ingoing into
the black hole, and the radiation is the future of the Schwarzschild region.

In order for the final region W to be a static-wormhole region, the mass
function $m$ of the Vaidya region V must take the following form;
\begin{equation}
\label{eq:m_{II}(z)}
m(z)=\frac{a}{2}(e^{-z^2}+2z\phi(z) -2\phi(z)^2e^{z^2})+C\phi(z) e^{z^2},
\end{equation}
and  the relation between the coordinates $V$ and $z$ is
\begin{equation}
\label{}
V(z)=\int _{0}^z\frac{2a^2e^{-y^2}(e^{-y^2}+2y\phi(y))}
{a\phi(y)-C}dy+V_0.
\end{equation}
From Eqs. (\ref{eq:m_{II}(V=V_0)}) and (\ref{eq:m_{II}(z)}),
$a$ is
\begin{equation}
a=2(M+\mu).
\end{equation}
Connecting the impulsive radiation (\ref{ET-sv}) and (\ref{ET-wv})
at $l=0$, the constant $C$ is decided as
\begin{equation}
C=\frac{\mu}{\sqrt{\lambda}}.
\end{equation}

\subsection{Wormhole region W}
We consider the spacetime in the region W.
Using the  matching formula (\ref{matchVW}) between Vaidya
(\ref{metric:Vaidya}) with the mass function (\ref{eq:m_{II}(z)}) and a
wormhole, we find that the region W is a static wormhole with the mass-energy
\begin{equation}
\epsilon(l)=(M+\mu)(e^{-l^2}+2l\phi(l) -2\phi(l)^2e^{l^2}).
\end{equation}
The relation between the throat radius $r_0$ of the wormhole in the final
region W and the Schwarzschild mass $M$ in the region S is
\begin{equation}
\label{jump0} r_0=a=2M+2\mu.
\end{equation}

Now we consider the energy and timing of the impulse.
The tortoise coordinate $r^{\ast}$ inside a Schwarzschild black hole can be
defined as
\begin{equation}
\label{tortoise}
r^{\ast}=-r-2M\ln\left(1-\frac{r}{2M}\right) \qquad {\rm for}\quad r<2M,
\end{equation}
so that $dr^{\ast}/dr>0$. In addition, the symmetry of the impulses means that
the intersection point O is given by $t = 0$, $r=r_0$ or $r^{\ast}=r^{\ast}_0$.
Then the Eddington-Finkelstein relation (\ref{EF}) at the point O where the
impulses collide gives
\begin{equation}
\label{r_1ast-mu}
V_0=-r^{\ast}_{0}=2(M+\mu)+2M\ln\left(\frac{-\mu}{M}\right).
\end{equation}
Thus the energy and timing of the impulses are related. From this relation,
first, the energy of the impulses must be always negative, $\mu<0$. The throat
radius of the final wormhole (\ref{jump0}) must be less than the horizon radius
of the initial Schwarzschild black hole. Second, the later the negative-energy
impulses occur, the larger the absolute value of the energy of the impulses
must be. These features are consistent with the results of the 2D model
\cite{KHK}.

In order for the final state not to have a naked singularity but to be a
wormhole, the inequality $-M<\mu<0$ is required. The throat radius of the
wormhole in the final region W must be smaller than the horizon radius of the
initial black hole, $r_0<2M$, since $\mu$ must be negative. In summary, one can
prescribe the initial black-hole mass $M>0$ and the impulse energy
$\mu\in(-M,0)$ as free parameters, then the timing $V_0$ of the impulses and
the throat radius $r_0$ of the final wormhole are determined.

\section{wormhole enlargement by impulsive radiation}
\label{sec:enlarge} 
\begin{figure}[t]
\begin{center}
\includegraphics[width=10cm]{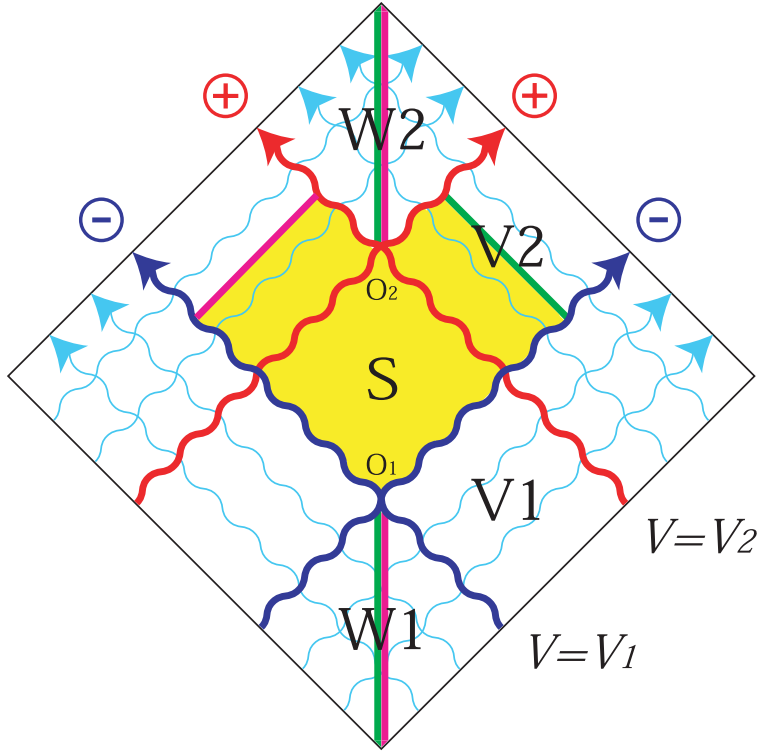}
\caption{(Color online). Penrose diagram of the wormhole enlargement model. The
wavy blue and red bold lines represent impulsive radiation with negative and
positive energy density, respectively. The regions W1 and W2 are
static-wormhole, V1 and V2 are Vaidya, and S is Schwarzschild. The boundary
between W1 and V1 corresponds to $z=l$, the boundary between V1 and S
corresponds to $z=0$, the boundary between S and V2 corresponds to $w=0$, and
the boundary between V2 and W2 corresponds to $w=l$.} \label{fig:enlarge}
\end{center}
\end{figure}

In this section we present an analytic solution which
represents the enlargement of a static wormhole. The whole picture is
represented by Fig. \ref{fig:enlarge} and the strategy is as follows. Basically
we want to open then close an expanding region of past trapped surfaces, by
moving apart then rejoining the two trapping horizons comprising the wormhole
throat in W1. The general recipe is to first strengthen then weaken the
negative energy density \cite{wh}. This can be done with a two-shot combination
of primary impulses with negative energy, followed by secondary impulses with
positive energy. To make the situation analytically tractable, the
constant-profile phantom radiation is turned off between the impulses, leaving
the region S as Schwarzschild and the regions V1, V2 as Vaidya. By controlling
the energy and timing of the impulses and the energy density of the final
constant-profile radiation, the final region W2 is also a static wormhole, but
larger.

\subsection{Vaidya region V1}
We set up the initial region W1 as a static wormhole (\ref{metric:wh}) with
throat radius $r_1$. Then the gravitational energy $\epsilon_{1}$ is
\begin{equation}
\epsilon_{1}(l)=\frac{r_1}{2}(e^{-l^2}+2l\phi(l) -2\phi(l)^2e^{l^2}).
\end{equation}
We beam in primary impulses symmetrically from both universes, then turn off
the constant ghost radiation immediately after the impulses. Then the region V1
should be Vaidya. Timing the impulses at $V=V_1$, the matching formula
(\ref{matchVW}) between static-wormhole and Vaidya regions
yields the mass-energy $m_{1}$ in the region V1 as
\begin{equation}
m_{1}(z)=\frac{r_1}{2}(e^{-z^2}+2z\phi(z) -2\phi(z)^2e^{z^2})+C_1\phi(z)
e^{z^2}
\end{equation}
where the relation between the coordinates $V$ and $z$ is given by
\begin{equation}
\label{}
V(z)=\int _{0}^z\frac{2r_1^2e^{-y^2}(e^{-y^2}+2y\phi(y))}
{r_1\phi(y)-C_1}dy+V_1.
\end{equation}
Connecting the impulsive radiation from the boundary between W and V1 to that
between V1 and S, we can decide the constant $C_1=\mu_1/\sqrt{\lambda}$, where
$\mu_1$ is the mass-energy of the primary impulses. Then the mass function of
the first Vaidya region V1 at the boundary with Schwarzschild S becomes
\begin{equation}
m_{1}(z=0)=\frac{r_1}{2}
\end{equation}
since the boundary surface $V=V_1$ coincides with $z=0$.

\subsection{Schwarzschild region S}
The region S is vacuum and therefore Schwarzschild.
From the matching formula between Vaidya and Schwarzschild (\ref{matchVS}), the
mass $M$ of the Schwarzschild region S becomes
\begin{equation}
\label{eq:M_III}
M=\frac{r_1}{2}+\eta_1\zeta_1\mu_1=\frac{r_1}{2}-\zeta_1\mu_1,
\end{equation}
where $\eta_1=-1$, since the Vaidya region V1 is to the past of the
Schwarzschild region S.
Since we construct solvable symmetric models, the impulses are also both
ingoing or outgoing. This means the region S must be inside a black-hole or
white-hole region and $r_1<2M$. So when the impulse has  negative energy,
$\mu_1<0$, the sign of $\zeta_1$ must be positive,
\begin{equation}
\label{jump1} M=\frac{r_1}{2}-\mu_1.
\end{equation}
This means the radiation is outgoing,
\begin{equation}
\frac{dr}{dV}>0,
\end{equation}
and S is a white-hole region. This is what we need to enlarge the wormhole, as
a white-hole region is expanding. Conversely, one could reduce the wormhole
size by taking $\mu_1>0$, creating a contracting black-hole region.

Now the symmetry of the impulses means that the intersection point $\rm O_1$ is
given by $t = 0$. Then the Eddington-Finkelstein relation (\ref{EF}) at the
point $\rm O_1$ where the impulses collide gives
\begin{equation}
\label{eq:r_1ast}
V_{1}=-r^{\ast}_1
=r_1+2M\ln\left(1-\frac{r_1}{2M}\right)
=2(M+\mu_1)+2M\ln\left(\frac{-\mu_1}{M}\right).
\end{equation}
Eqs.~(\ref{eq:M_III}) and (\ref{eq:r_1ast}) mean that the timing $V_0$ of the impulses and the Schwarzschild mass $M$ are determined by the throat radius $r_1$ of the initial wormhole and the energy $\mu_1$ of the impulses.

\subsection{Vaidya region V2}
We next beam in secondary impulses symmetrically from both universes, and turn
on constant-profile phantom radiation immediately after the impulses. Then the
region V2 must be Vaidya. Timing the impulses at $V=V_2$, the matching formula
between Vaidya and Schwarzschild (\ref{matchVS}) yields the mass function of
the second Vaidya region V2 as
\begin{equation}
m_{2}(V_2)=M-\eta_2\zeta_2\mu_2
\end{equation}
on the boundary, where $\mu_2$ is the mass of the second shell. The region V2
is an outgoing Vaidya region which is to the future of the Schwarzschild
region, so that $\eta_2 =1$ and $\zeta_2=1$. Then
\begin{equation}
m_{2}(V_2)=M-\mu_2=\frac{r_1}{2}-\mu_1-\mu_2.
\end{equation}
Here in order for the final
region W2 to be a static wormhole, the mass function $m_{2}$ must take the
following form,
\begin{equation}
\label{eq:m_{V}(w)}
m_{2}(w)=\frac{1}{2}(r_1-2\mu_1-2\mu_2)
(e^{-w^2}+2w\phi(w) -2\phi(w)^2e^{w^2})
+C_2\phi(w) e^{w^2},
\end{equation}
where the coordinate $w$ is related with $V$ by
\begin{equation}
\label{}
V(w)=\int _{0}^w\frac{2r_2^2e^{-y^2}(e^{-y^2}+2y\phi(y))}
{r_2\phi(y)-C_2}dy+V_2,
\end{equation}
from the matching formula (\ref{eq:M_V}). Here $r_2$ is the throat radius of
the wormhole in the final region W2. Connecting the impulsive radiation from
the boundary between the regions S and V2 to that between V2 and W2, we can
decide the constant $C_2=\mu_2/\sqrt{\lambda}$.

It can be shown that there are trapping horizons in the regions V2, as depicted
in Fig \ref{fig:enlarge}; the negative-energy and positive-energy impulses
respectively make the horizons jump to the future and the past. It is difficult
to study the horizons analytically in the Vaidya coordinates, but it can be
shown that they are null using dual-null coordinates $x^\pm$, as follows. If we
have $V$ pointing along $x^+$, then there is only a $T_{++}$ component in the
energy tensor. The $T_{--}$ and $T_{+-}$ components of the Einstein equations
\cite{bhd} then show respectively that, where $\partial_-r=0$, then
$\partial_-\partial_-r=0$ and $\partial_+\partial_-r<0$, which means that the
horizon $\partial_-r=0$ is null. Similar behavior occurs in the 2D model
\cite{KHK}, though the horizons were omitted in the corresponding diagram.

\subsection{Wormhole region W2}
Finally, we consider the spacetime in the region W2. Since constant-profile
phantom radiation is beamed in for $V>V_2$ in order for the region V2 to be
Vaidya (\ref{metric:Vaidya}) with the mass function (\ref{eq:m_{V}(w)}), we can
match it to a static-wormhole region from the matching formula (\ref{matchVW}).
We find that the mass function $\epsilon_{2}$ of the wormhole in the region W2
is
\begin{equation}
\epsilon_{2}(l)=(r_1-2\mu_1-2\mu_2)(e^{-l^2}+2l\phi(l) -2\phi(l)^2e^{l^2}),
\end{equation}
and the throat radius $r_2$ is
\begin{equation}
\label{jump2} r_2=r_1-2\mu_1-2\mu_2=2M-2\mu_2.
\end{equation}

Again, the symmetry of the impulses means that the intersection point $\rm O_2$
is given by $t = 0$. Then the Eddington-Finkelstein relation (\ref{EF}) at the
point $\rm O_2$ where the impulses collide gives
\begin{equation}
\label{eq:r_2ast}
V_2=-r^{\ast}_2
=r_2+2M\ln\left(1-\frac{r_2}{2M}\right)
=2(M+\mu_2)+2M\ln\left(\frac{\mu_2}{M}\right).
\end{equation}
From this relation, the energy of the impulses must be positive, $\mu_2>0$. In
addition, the inequality
\begin{equation}
\label{ineq:r_2-r_1}
r_{2}>r_{1}
\end{equation}
holds, since the region S is part of a white-hole region. That is, the wormhole
is enlarged. We find that the absolute value $|\mu_1|$ of the energy density of
the primary impulses should be stronger than that of the secondary impulses,
\begin{equation}
|\mu_1|>|\mu_2|
\end{equation}
from Eqs.~(\ref{jump2}) and (\ref{ineq:r_2-r_1}).

We find the relation
between the energy and timing of impulses as
\begin{equation}
\label{v_2-v_1}
V_2-V_1=r^{\ast}_2-r^{\ast}_1
=2(\mu_1+\mu_2)+2M\ln\left(\frac{\mu_2}{-\mu_1}\right),
\end{equation}
from Eqs. (\ref{eq:r_1ast}) and (\ref{eq:r_2ast}). Eq. (\ref{v_2-v_1}) means
that the longer the interval between the first and second impulses is, the
smaller the value of energy density of the second impulse must be. In summary,
once the throat radius of the initial wormhole and the energy of the impulses
$(r_1,\mu_1,\mu_2)$ are prescribed, the timings of the impulses, the
intermediate Schwarzschild mass and the throat radius of the final wormhole
$(V_1,M,V_2,r_2)$ are determined. These features are also consistent with the
results of the 2D model \cite{KHK}.

\section{jump in energy due to impulsive radiation}
\label{sec:jump} A general spherically symmetric metric can be written in
dual-null form as
\begin{equation}
 ds^2=r^2d\Omega^2-h\,dx^+dx^-
\end{equation}
where $r\ge0$ and $h>0$ are functions of the future-pointing null coordinates
$(x^+,x^-)$. Writing $\partial_\pm=\partial/\partial x^\pm$, the propagation
equations for the energy $E$ (\ref{energy}) are obtained from the Einstein
equations as \cite{1st}
\begin{equation}
\partial_\pm E=8\pi h^{-1}r^2(T_{+-}\partial_\pm r-T_{\pm\pm}\partial_\mp r).
\end{equation}
We have considered impulsive radiation defined by
\begin{equation}
T_{ab}=\frac{\mu_\pm\delta_a^\pm\delta_b^\pm}{4\pi r^2}\delta(x^\pm-x_0)
\end{equation}
where the constant $x_0$ gives the location of the impulse and the constant
$\mu_\pm$ is its energy. More invariantly, the vector $\wp=-g^{-1}(\mu_\pm
dx^\pm)$ is the energy-momentum of the impulse. Then the jump
\begin{equation}
[E]_\pm=\lim_{\alpha\to0}\int_{x_0-\alpha}^{x_0+\alpha}\partial_\pm E\,dx^\pm
\end{equation}
in energy across the impulse is given by the jump formula
\begin{equation}
[E]_\pm=c^\pm\mu_\pm,\qquad c^\pm=-2(h^{-1}\partial_\mp r)\vert_{x^\pm=x_0}.
\end{equation}
The vector $c=c^+\partial_++c^-\partial_-$ is actually $c=g^{-1}(dr)$ and so
\begin{equation}
[E]_\pm=-\wp\cdot dr
\end{equation}
is a manifestly invariant form of the jump formula. Note that while the
energy-momentum vector $\wp$ (or $\mu_\pm dx^\pm$) is invariant, the energy
$\mu_\pm$ depends on the choice of null coordinate $x^\pm$, reflecting the fact
that a particle moving at light-speed has no rest frame and no preferred
energy. However, in a curved but stationary space-time, the stationary Killing
vector provides a preferred frame and a preferred energy $\mu_\pm$.

We need only employ the jump formula in the following cases. (i) Inside a
Schwarzschild black hole, we can take future-pointing $x^\pm=-r^{\ast}\pm t$
where $dr^{\ast}/dr=(2M/r-1)^{-1}$. Then $r^{\ast}=-(x^++x^-)/2$, $h=2M/r-1$
and $\partial_\pm r=(dr/dr^{\ast})\partial_\pm r^{\ast}=(1-2M/r)/2$ gives
$c^\pm=1$ and $[E]_\pm=\mu_\pm$. (ii) Inside a Schwarzschild white hole, we can
take future-pointing $x^\pm=r^{\ast}\pm t$ and similarly obtain $c^\pm=-1$ and
$[E]_\pm=-\mu_\pm$. (iii) On the throat of a static wormhole, where
$\partial_+r=\partial_-r=0$ and $h$ is finite \cite{MT}, one finds $c^\pm=0$
and $[E]_\pm=0$. This is summarized as
\begin{equation}
[E]_\pm=\left\{\begin{array}{ll} \mu&\quad\hbox{inside a
Schwarzschild black hole}\\
-\mu&\quad\hbox{inside a Schwarzschild white hole}\\
0&\quad\hbox{on the throat of a static wormhole}
\end{array}
\right.
\end{equation}
where the indices on $\mu_\pm$ are now omitted.

\begin{figure}[h]
\includegraphics[width=7cm]{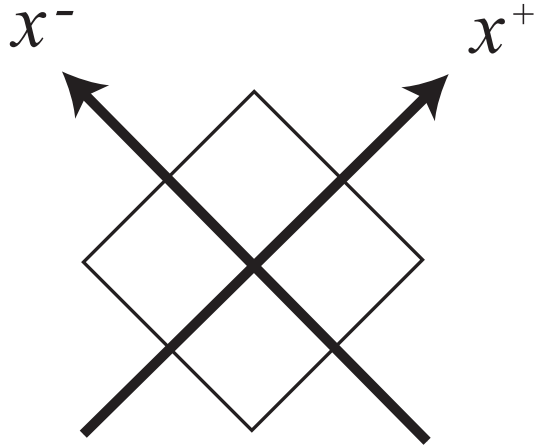}
\caption{an infinitesimal box across which two radiative impulses (arrows)
intersect.} \label{fig:intersect}
\end{figure}

Now assuming an infinitesimal diamond-shaped box around the point where the
impulses collide as in Fig.\ref{fig:intersect}, we can evaluate the jump in
energy across the impulses for each case in Secs.~IV and V. For wormhole
construction (Fig.\ref{fig:sudden}), the energy $E$ will jump by $\mu$ from the
region S to V and by $0$ from the region V to W, evaluated in the limit at the
point, recovering $r_0/2=M+\mu$ (\ref{jump0}), where $M$ is the black-hole mass
in the region S. Similarly for wormhole enlargement (Fig.\ref{fig:enlarge}),
the energy $E$ will jump by $0$ from W1 to V1 and by $-\mu_1$ from V1 to S at
the point $\rm O_1$, and by $-\mu_2$ from the region S to V2 and by $0$  from
the region V2 to W2 at $\rm O_2$, recovering $M=r_1/2-\mu_1$ (\ref{jump1}) and
$r_2/2=M-\mu_2$ (\ref{jump2}), where $M$ is the black-hole mass in the region
S.

Thus even without performing the detailed matching, the basic properties of the
solutions could be predicted simply by the jump formula for $E$ and continuity
of the area $A=4\pi r^2$. For wormhole construction, continuity of $r$ at O
implies $r_0<2M$, and the jump formula gives $\mu<0$; the impulses must have
negative energy. Similarly, for wormhole enlargement, continuity of $r$ at O1
and O2 implies $r_1<2M$ and $r_2<2M$, and the jump formula gives $\mu_1<0$ and
$\mu_2>0$; the primary and secondary impulses must have negative and positive
energy respectively.


\section{Summary}
In this paper, we have studied wormhole dynamics in Einstein gravity under
(phantom and normal, impulsive and regular) pure radiation, constructing
analytic solutions where a traversable wormhole is created from a black hole,
or the throat area of a traversable wormhole is enlarged or reduced, the size
being controlled by the energy and timing of the impulses. The solutions are
composed of Schwarzschild, static-wormhole \cite{pr} and Vaidya regions matched
across null boundaries according to the Barrab\`es-Israel formalism. For this
purpose we have derived the matching formulas which apply when the direction of
radiation in the Vaidya region is either parallel or transverse to the
boundary. These formulas are useful for other problems.

The results provide concrete examples of how to create and enlarge traversable
wormholes, given the existence of Schwarzschild black holes and phantom energy.
We have worked within standard General Relativity, inventing no new theoretical
physics other than an idealized model of phantom energy, on which general
arguments do not depend \cite{wh}. Thus if space-time foam and phantom energy
do exist and can be controlled, then traversable wormholes can be constructed
and enlarged.

\acknowledgements H.K. is supported by JSPS Research Fellowships for Young
Scientists.


\begin{references}
\bibitem{MT} M.S. Morris \& K.S. Thorne, Am. J. Phys. {\bf 56}, 395
 (1988).
\bibitem{V}M. Visser, Lorentzian Wormholes: from Einstein to Hawking (AIP
 Press 1995).
\bibitem{Spe}D.N. Spergel et al., {Astroph. J. Suppl.} {\bf 148}, 175 (2003).
\bibitem{Kra}L.M. Krauss, {Astroph. J.} {\bf 596}, L1 (2003).
\bibitem{Cal}R.R. Caldwell, {Phys. Lett.} {\bf B545}, 23 (2002).
\bibitem{HV1}D. Hochberg \& M. Visser, {Phys. Rev.} {\bf D58}, 044021 (1998).
\bibitem{HV2}D. Hochberg \& M. Visser, {Phys. Rev. Lett.} {\bf 81}, 746 (1998).
\bibitem{wh} S.A. Hayward, {Int. J. Mod. Phys.} {\bf D8}, 373 (1999).
\bibitem{IH}D. Ida \&  S. A. Hayward, {Phys. Lett.} {\bf A260}, 175 (1999).
\bibitem{bhd}S.A. Hayward, {Phys. Rev.} {\bf D49}, 6467 (1994).
\bibitem{1st}S.A. Hayward, {Class. Quantum Grav.} {\bf 15}, 3147 (1998).
\bibitem{HKL} S.A. Hayward, S-W. Kim \& H. Lee,
Phys. Rev. {\bf D65}, 064003 (2002).
\bibitem{SH}H. Shinkai \& S.A. Hayward, {Phys. Rev.} {\bf D66}, 044005
 (2002).
\bibitem{pr} S.A. Hayward, Phys. Rev. {\bf D65}, 124016 (2002).
\bibitem{KHK}H. Koyama, S.A. Hayward \& S-W. Kim, {Phys. Rev.} {\bf D67},
 084008 (2003).
\bibitem{B-I} C. Barrab\`es \& W. Israel, Phys. Rev. {\bf D43}, 1129 (1991).
\bibitem{Vaidya}
P.C. Vaidya, {\it Proc. Indian Acad. Sci., Sect.} {\bf A33}, 264 (1951).
\bibitem{Whe}J.A. Wheeler, {Ann. Phys.} {\bf 2}, 604 (1957).
\bibitem{HK}S.A. Hayward \& H. Koyama,
How to make a traversable wormhole from a Schwarzschild black hole
(gr-qc/0406080).
\bibitem{MS}C.W. Misner \& D.H. Sharp, {Phys. Rev.} {\bf 136}, B571 (1964).
\bibitem{sph}S.A. Hayward, {Phys. Rev.} {\bf D53}, 1938 (1996).
\bibitem{Ellis} H.G. Ellis, J. Math. Phys. {\bf 14}, 104 (1973).
\bibitem{Gergely}L.A. Gergely, {Phys. Rev.} {\bf D58}, 084030 (1998).
\bibitem{Jez} J. Jezierski, J. Math. Phys. {\bf 44}, 641 (2003).
\end{references}
\end{document}